\documentclass[runningheads]{llncs}
\usepackage{graphicx}

\usepackage[utf8]{inputenc}
\usepackage{color}
\usepackage{soul}

\usepackage{hyperref}
\usepackage{wrapfig}
\usepackage{amssymb}
\usepackage{array, makecell}
\usepackage{tikz}                  \usetikzlibrary{calc,arrows,positioning,shapes,shadows,spy,plotmarks,matrix,fit,backgrounds}

\begin{document}
\title{Hybrid Cascaded Neural Network for \\ Liver Lesion Segmentation}
%
%
\author{Raunak Dey, Yi Hong}
\authorrunning{ }
%
\institute{Computer Science Department, University of Georgia}
\maketitle              

\begin{abstract}
Automatic liver lesion segmentation is a challenging task while having a significant impact on assisting medical professionals in the designing of effective treatment and planning proper care. In this paper we propose a cascaded system that combines both 2D and 3D convolutional neural networks to effectively segment hepatic lesions. Our 2D network operates on a slice by slice basis to segment the liver and larger tumors, while we use a 3D network to detect small lesions that are often missed in a 2D segmentation design. We employ this algorithm on the LiTS challenge obtaining a Dice score per case of $68.1\%$, which performs the best among all non pre-trained models and the second best among published methods. We also perform two-fold cross-validation to reveal the over- and under-segmentation issues in the LiTS annotations.

\keywords{Liver lesion segmentation \and Hybrid neural network \and Small lesion segmentation.}
\end{abstract}

\section{Introduction}
Liver lesions are groups of abnormal cells in the liver and some of them lead to the cancer. Liver cancer is one of the leading causes of cancer deaths worldwide and more than 700,000 deaths are reported each year according to the American Cancer Society. For the liver cancer screening, the Computer Tomography (CT) is the most commonly used imaging tool and the technique of the automatic liver lesion segmentation from a CT scan has great impacts on cancer diagnosis, surgery planning, and treatment evaluations.

Due to the heterogeneous and diffusive appearance of hepatic lesions, the liver lesion segmentation is a challenging task. Researchers have proposed many segmentation algorithms based on the classical segmentation techniques, e.g., thresholding~\cite{moltz2008segmentation}, region growing~\cite{moltz2009advanced}, active contour~\cite{massoptier2008new}, and ensemble~\cite{huang2014random}. Recently, deep neural networks have been widely used in the liver lesion segmentation and have shown improved performance in segmenting and detecting liver lesions. These algorithms either use 2D convolutional neural networks~\cite{chlebus2018automatic,han2017automatic,li2018bottleneck}, 3D networks~\cite{jiang2019ahcnet}, or a combination of both~\cite{christ2017automatic,liu20183d}. One of the issues facing by the existing methods is the segmentation of the small lesions, which are often missed in their segmentation predictions. 




In this paper we introduce a hybrid cascaded segmentation network to segment the liver lesions, especially the small ones. The network utilizes both 2D and 3D convolutional networks to 
effectively find both large and small lesions in the liver region.  We use a 2D convolutional neural network (CNN) to obtain the liver mask from the input CT volume of the abdomen region, which locates the liver region for further processing. Another 2D CNN is then employed to extract large lesions from 2D slices, while a dedicated 3D CNN is proposed to segment out small lesions from 3D volumes. 

We evaluate our method on the Liver Tumor Segmentation (LiTS) dataset~\cite{bilic2019liver} and obtain a Dice score per subject of $68.1\%$ without any pre-training or post-processing. This result ranks first among the currently published non pre-trained networks. Moreover, with the proposed special treatment of the small lesion segmentation using a 3D CNN, we improve the network performance of segmenting the large lesions only by $7.1\%$ in terms of the Dice score per subject. In addition, using a two-fold cross-validation on the LiTS training set, we observe that our network can segment unannotated regions which share the similar intensity as that of the tumor region annotated on the same slice in the LiTS dataset.







\begin{figure}[!t]
\centering
\includegraphics[width=1.0\textwidth]{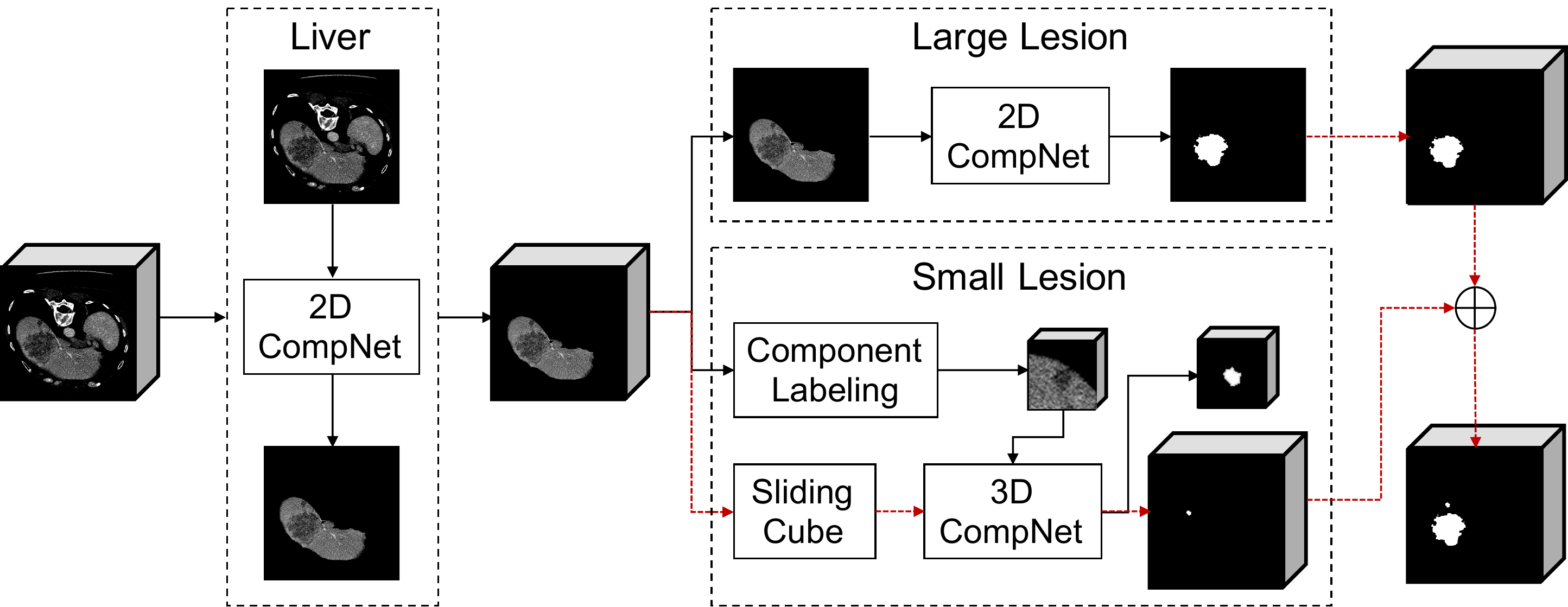}
\caption{Overview of our proposed hybrid cascaded network for the liver lesion segmentation. We adopt 2D CompNets (see Fig.~\ref{fig:liverSeg} for architecture details) to segment the liver region and large lesions and apply a 3D CompNet (Fig.~\ref{fig:small_tumor_Seg}) to segment the small lesions. The red dashed lines indicate the test phase of the segmentation.}
\label{fig:overview}
\end{figure}


\section{Method}

\subsection{Overview}
In this section we describe our hybrid cascaded network to handle the liver lesion segmentation problem. We first segment the liver from the original CT scan slice-by-slice using a 2D neural network. The generated liver mask allows us to focus on the lesion segmentation located inside of the liver. At the stage of lesion segmentation, we treat large and small lesions separately with a 2D network branch to segment large lesions and a 3D network branch to segment small lesions. This hybrid design was motivated by the goal of balancing the computational efficiency without sacrificing the accuracy. Compared to the 3D segmentation network, a 2D network is more efficient in terms of both computation time and memory cost; however, on a 2D slice small lesions are likely to be confused with some normal tissues in the liver region. Therefore, we propose to use an adaptive 3D network to handle the small lesion segmentation, which greatly reduces such false positives. Figure~\ref{fig:overview} depicts the overview of the proposed hybrid cascaded network. Its backbone network is the CompNet~\cite{dey2018compnet}, which was demonstrated to be more robust in the segmentation on pathological images than the U-Net~\cite{ronneberger2015u}.    



\begin{figure}[t]
\centering
\includegraphics[width=1.0\textwidth]{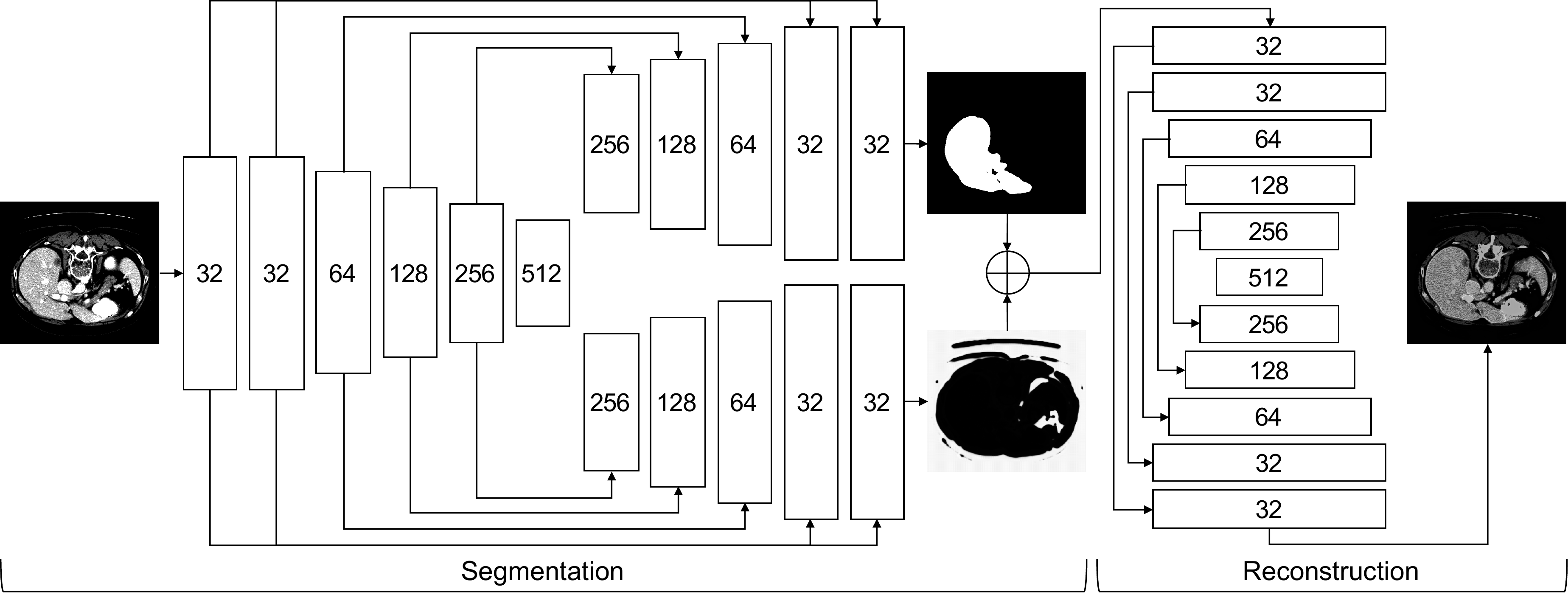}
\caption{Network architecture of the 2D CompNet for the liver/large lesion segmentation. This U-Net-like network has the object segmentation (top left), its complementary segmentation (bottom left), and the image reconstruction (right). Each block has two convolutional layers with the numbers of feature maps showing in the block box.}
\label{fig:liverSeg}
\end{figure}

\subsection{Liver Segmentation}
A liver CT scan typically has hundreds of slices and each slice has a resolution of $512 \times 512$ pixels. Directly working on such 3D images would cause time and memory issues using a segmentation neural network under an encoder-and-decoder design. Fortunately, the segmentation of the liver region inside the abdominal cavity is relatively easier than the segmentation of the liver lesions. As demonstrated by recent works~\cite{chlebus2018automatic,han2017automatic,li2018h,yuan2017hierarchical} and our experiments, liver segmentation relied only on 2D slices can achieve more than $95\%$ overall dice accuracy. Therefore, we use a 2D network to segment out the liver from a 3D CT scan. 

We follow the design of the basic CompNet proposed in~\cite{dey2018compnet} and adapt it to our liver segmentation task, resulting in the 2D network architecture shown in Fig.~\ref{fig:liverSeg}. This network takes 2D slices as input and outputs 2D liver masks that are stacked into a 3D mask. The encoder branches of the liver segmentation, its complementary segmentation, and the reconstruction sections consist of multiple blocks of two convolutional layers followed by the batch normalization layers. All convolutional layers use a filter size of $3\times 3$ and the filter numbers in the blocks are as follows 32, 32, 64, 128, 256, and the transition block of 512.
Following each block we have a pooling layer. The decoder branches of the network mirror the encoder ones with the pooling layers being replaced by 2D transposed convolutional layers. We multiply the input 3D scan with the obtained liver mask to retain only the liver region for the following lesion segmentation.



\begin{wrapfigure}{r}{0.4\textwidth}
\vspace{-0.3in}
\centering
\begin{tikzpicture}[thick, spy using outlines={circle,lens={scale=2.5}, width=2cm, height=1.5cm, connect spies}]
	\node (reg_id1) {\includegraphics[width=0.4\textwidth]{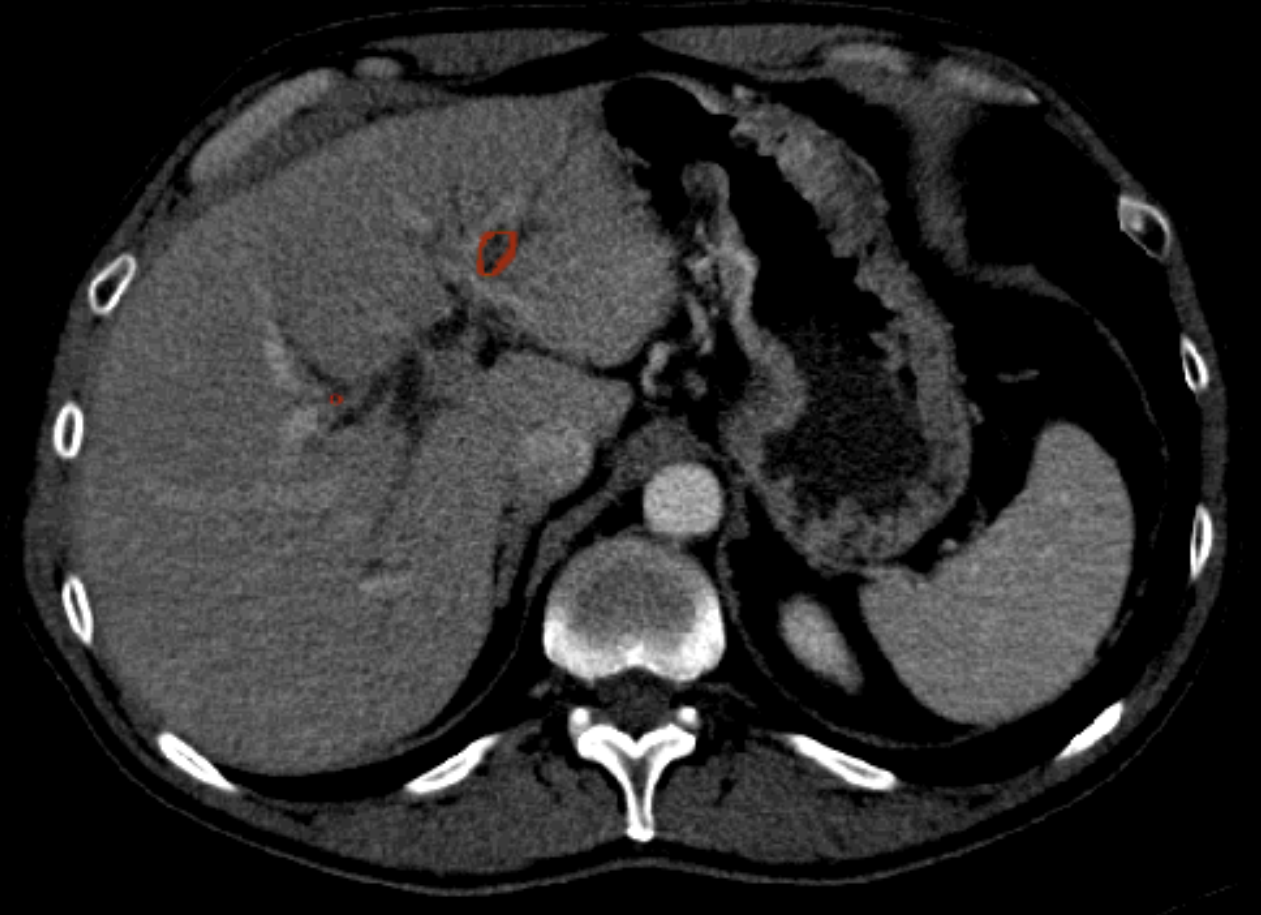}};
 	\spy [yellow] on (-0.5, 0.8) in node (lspy) [left,fill=black!5] at (2.4,1.2);
 	\spy [yellow] on (-1.1, 0.22) in node (lspy) [left,fill=black!5] at (-0.4,-1.2);
\end{tikzpicture}
\caption{False positive examples of small liver lesions on a 2D slice.}
\label{fig:lesion_false_positive}
\end{wrapfigure}

\subsection{Two-Step Lesion Segmentation}
After having the liver segmented, we might directly apply a 3D segmentation network to extract the liver lesions. However, a 3D Network operating only in the liver region is still computationally expensive and may suffer from the problem of lacking 3D training scans. In order to mitigate these issues, we propose to use a mixed network to handle the segmentation of large and small lesions separately. In particular, we use a 2D network to scan slice-by-slice and predict larger tumors if present. However, this method is not sufficient for predicting small lesions, as the false positive examples shown in Fig.~\ref{fig:lesion_false_positive}. This happens due to the appearance of the small lesion matching that of other tissues or vessels on 2D liver slices. A 3D network learning from cropped volumes with small lesions can reduce such false positives, because it leverages the observation that the location of the tissue or vessel appears to travel considerably in-between slices whereas the liver lesion constantly appears to be roughly stationary across slices. Regarding the threshold of the lesion size used to separate the large and small lesions, it may vary for different applications and datasets. Here, we set the threshold as a resolution of $32\times 32 \times 32$ by experimentally testing on the LiTS dataset, which greatly balances the prediction accuracy and the computational efficiency. 


\noindent
{\bf Large Lesion Segmentation.}
We use another 2D CompNet like the one used for the liver segmentation (Fig.~\ref{fig:liverSeg}) to extract lesions larger than $32\times 32$ from the input slices with the liver only. Since this network takes care of the prediction of large lesions, we clean the training masks by using the component labeling~\cite{opencv_library} to remove all small lesion masks, whose horizontal and vertical dimensions are both less than or equal to 32 on a 2D slice, for every slice. In the test phase we also remove the detected lesions less than $32 \times 32$ using the component labeling.

\begin{figure}[!t]
\centering
\includegraphics[width=0.78\textwidth]{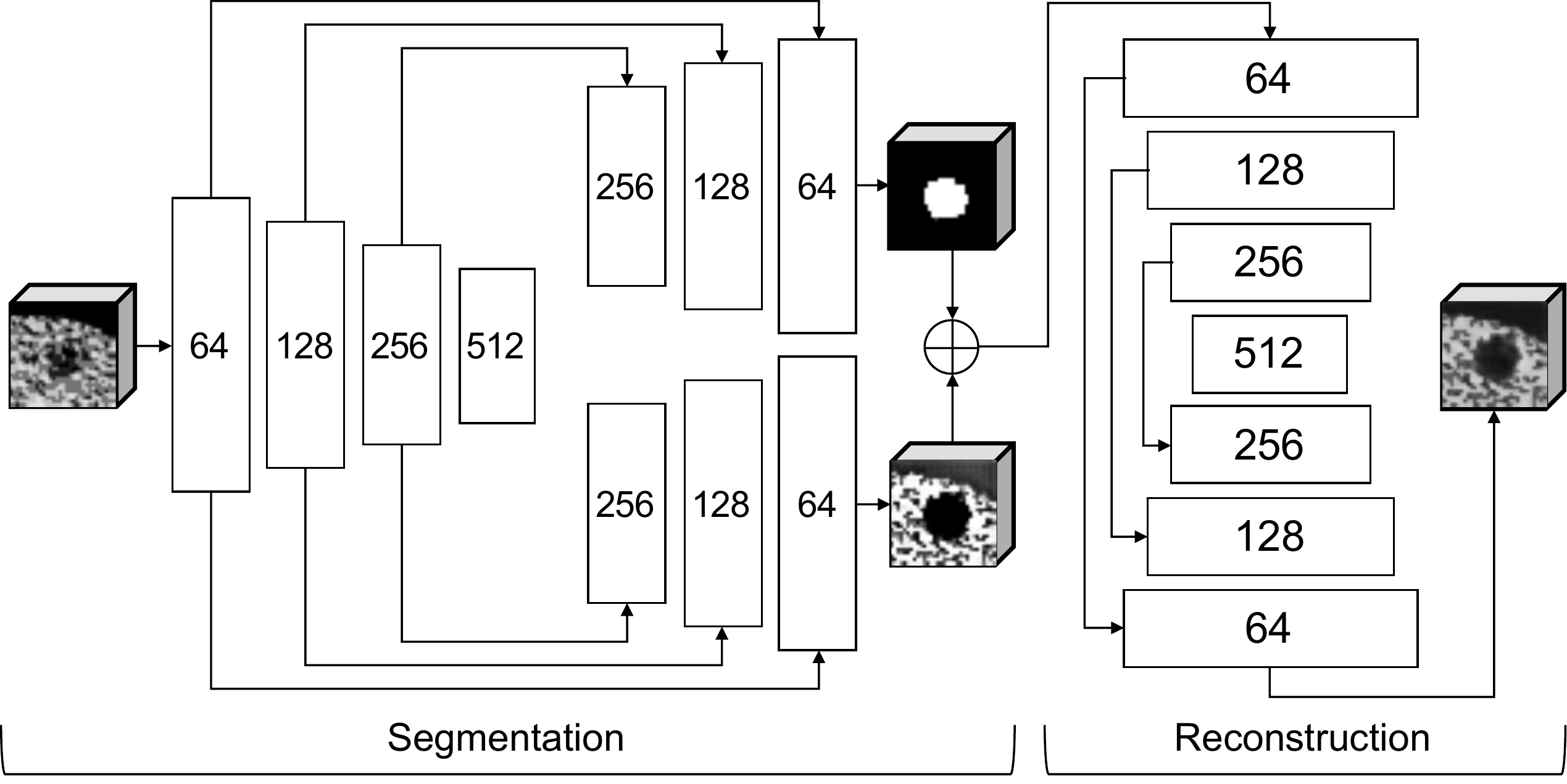}
\caption{Network architecture of the 3D CompNet used for the small lesion segmentation. This U-Net-like network has the small lesion segmentation (top left), its complementary segmentation (bottom left), and the image reconstruction (right). Each block has two convolutional layers with the numbers of feature maps showing in the block box.}
\label{fig:small_tumor_Seg}
\end{figure}

\noindent
{\bf Small Lesion Segmentation.}
We use a 3D CompNet as shown in Fig.~\ref{fig:small_tumor_Seg} to segment the small lesions. This 3D network takes a volume of $32 \times 32 \times 32$ as input. The encoder branches of the lesion segmentation, its complementary segmentation, and the reconstruction consist of three 3D convolutional blocks. Each block is composed of two 64, 128, and 256 feature maps, respectively, each followed by a 3D pooling layer with a pool size $(2,2,2)$. The transition block has 512 feature maps. The decoder branches mirror the encoder branches with the pooling layers replaced by the 3D transposed convolutional layers.

To generate the training samples for this 3D network, we use the component labeling to locate the center of a lesion and estimate its dimensions on a 2D slice. We sample a 3D volume for a lesion whose both horizontal and vertical dimensions on the 2D slice are smaller than or equal to $32$. Since the small lesions often occupy only a few slices, we choose a cube with a size of $32 \times 32 \times 32$ to cover the lesion. In particular, around the lesion center located on a slice, we select 15 slices above and 16 slices below to create a 3D volumes for training.


During the testing phase, we use a 3D sliding cube over the liver volume to predict the small lesions using the above trained 3D network. To choose an appropriate stride size for the sliding cube, we test on 32, 16, 8, and 4 voxels, respectively. It turns out that the stride 32 and 16 are too large to capture tiny lesions. The stride 8 is an appropriate one, since a further reduction of the stride size does not reveal any further changes but increasing the prediction time. To fuse the overlap predictions, we take the average of all predictions at that voxel and set the value greater than 0.5 to 1 as the final prediction. 




\section{Experiments}
\noindent
{\bf Dataset and Preprocessing.}
To test our proposed network, we use the public Liver Tumor Segmentation (LiTS)~\cite{bilic2019liver} dataset, which consists of 130 abdomen CT scans for training and 70 for testing. To train the 2D liver segmentation network, we fully use all training scans with a total of 58,638 2D slices. In the network training of the large liver lesion segmentation, we focus on the slices with the liver present, resulting in 19,163 2D slices in total. The 3D network is trained on 11,503 3D small lesion samples with size of $32 \times 32 \times 32$. 

%
%
%

\begin{wrapfigure}{r}{0.51\textwidth}
\centering
  \includegraphics[width=0.25\textwidth]{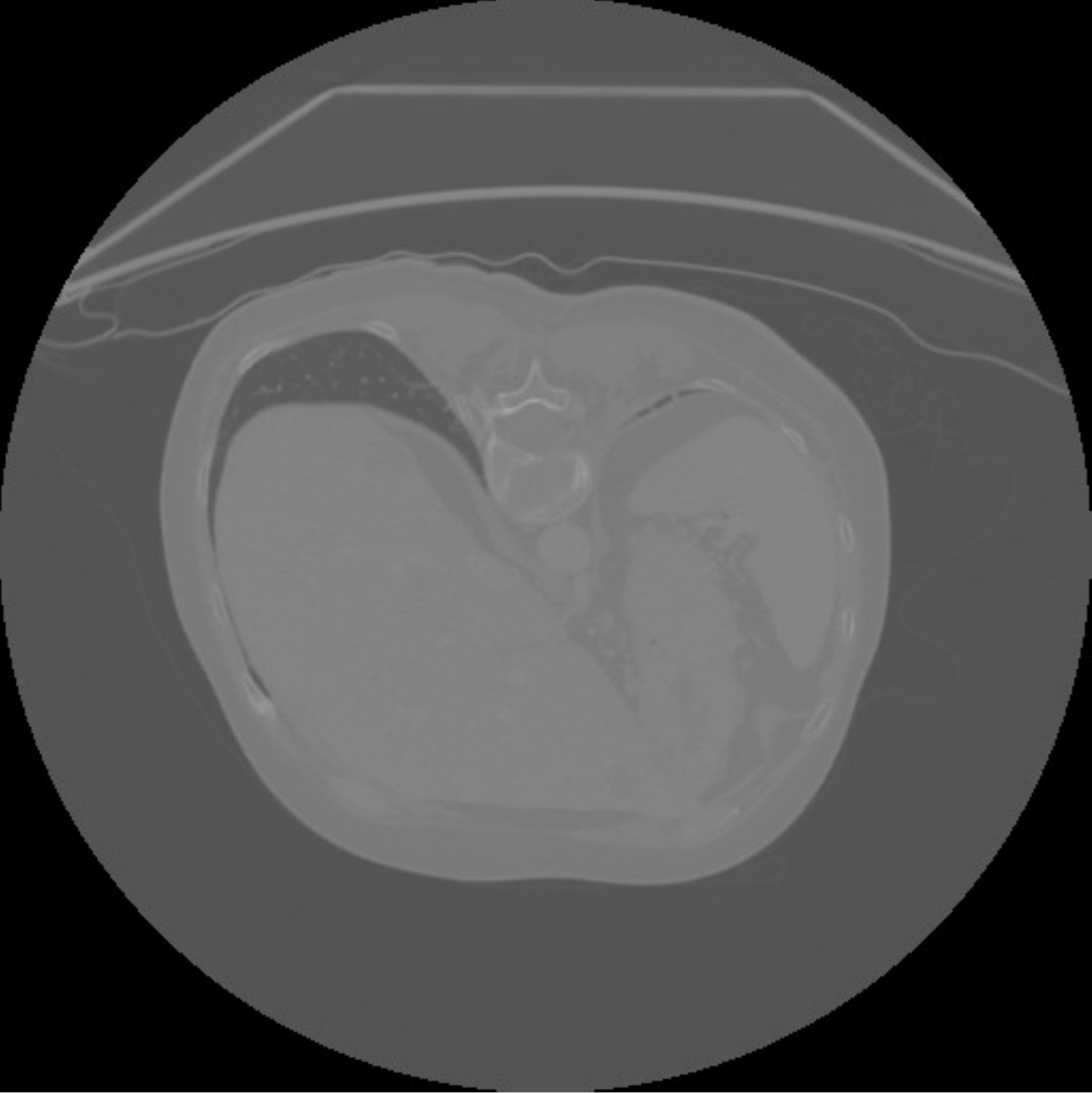}
  \includegraphics[width=0.25\textwidth]{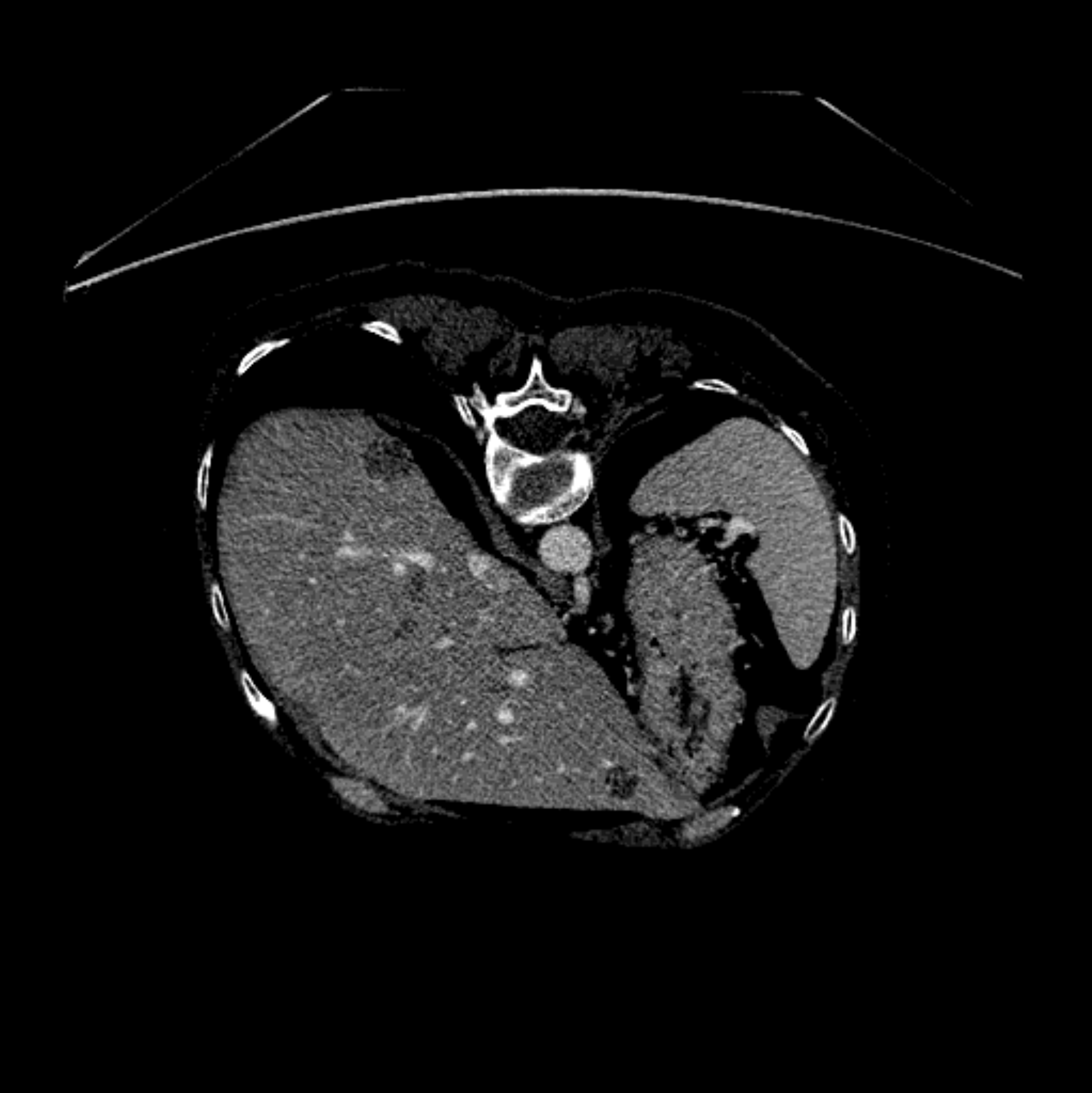}
\caption{An example of a liver CT scan before (left) and after (right) preprocessing.}
\label{fig:preprocessing}
\end{wrapfigure}

We preprocess the liver CT scans using a histogram based thresholding method. We select the rightmost peak of the intensity histogram distribution of a CT scan for normalization and use the histogram equalization algorithm to generate the enhanced images as shown in Fig.~\ref{fig:preprocessing}. {The implementation of the preprocessing and the liver lesion segmentation network is available online {\footnote{\url{https://github.com/raun1/LITS_Hybrid_Comp_Net}}}}.

\noindent
{\bf Experimental Settings.}
We use Keras with the TensorFlow backend to implement our proposed network. The 2D CompNet for the liver segmentation is trained for 40 epochs using the Adam optimizer with a learning rate of 5e-5. Next, we train the 2D and 3D CompNets for the large and small lesion segmentation in the same manner, i.e., we first train the networks using the Adam optimizer with a learning rate of 5e-5, the same as that used in \cite{chlebus2018automatic}, and having an early stopping scheme with the tolerance being set to 5; then we train the networks with a learning rate of 1e-6 using an early stopping with a tolerance of 10 trials. Both steps have 150 maximum number of epochs for training. 
In addition, we use an L2 regularization with a parameter of 2e-4 and a dropout with a rate of 0.3 after all pooling and upsampling layers to mitigate overfitting.

\begin{figure}[b]
\vspace{-0.2in}
\centering
  \includegraphics[height=0.3\textwidth]{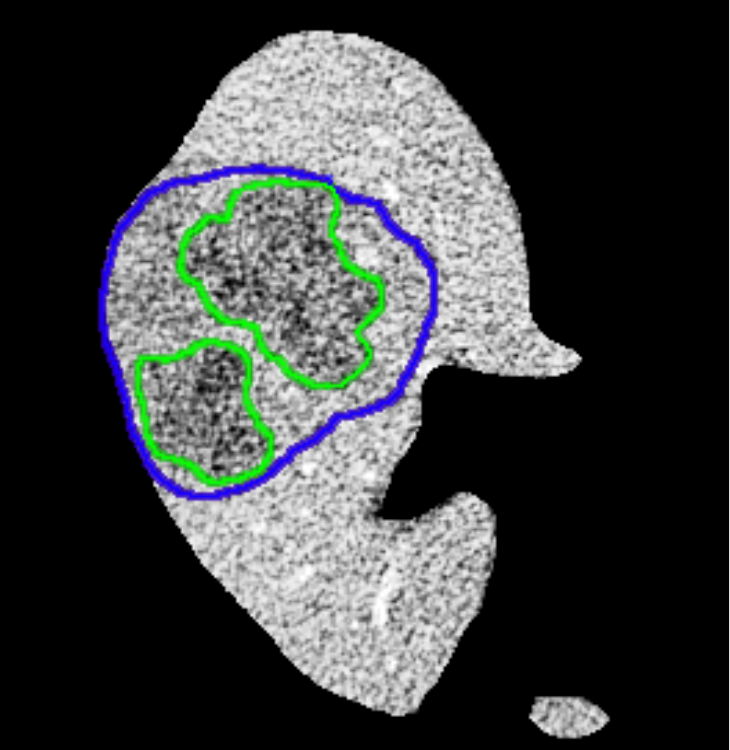}
  \includegraphics[height=0.3\textwidth]{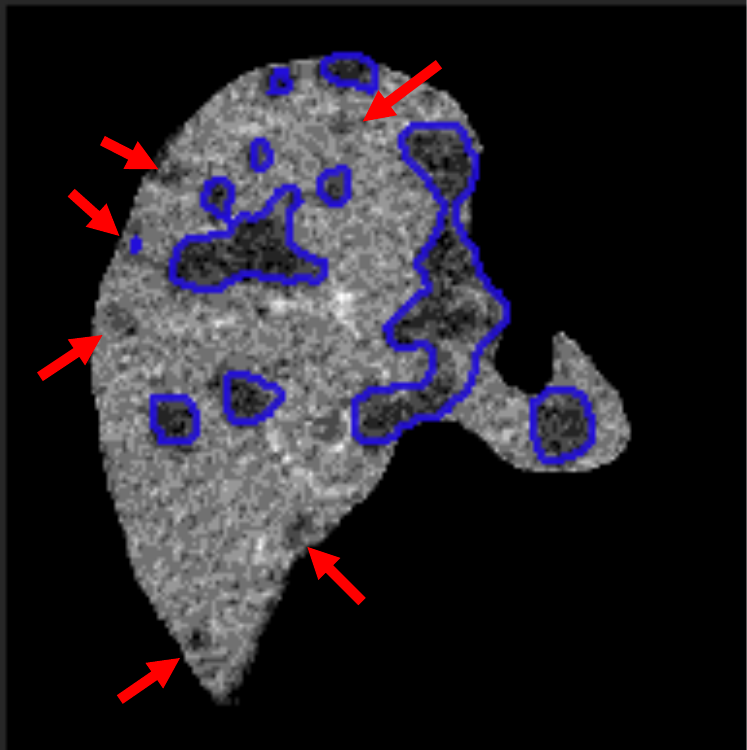}
  \includegraphics[height=0.3\textwidth]{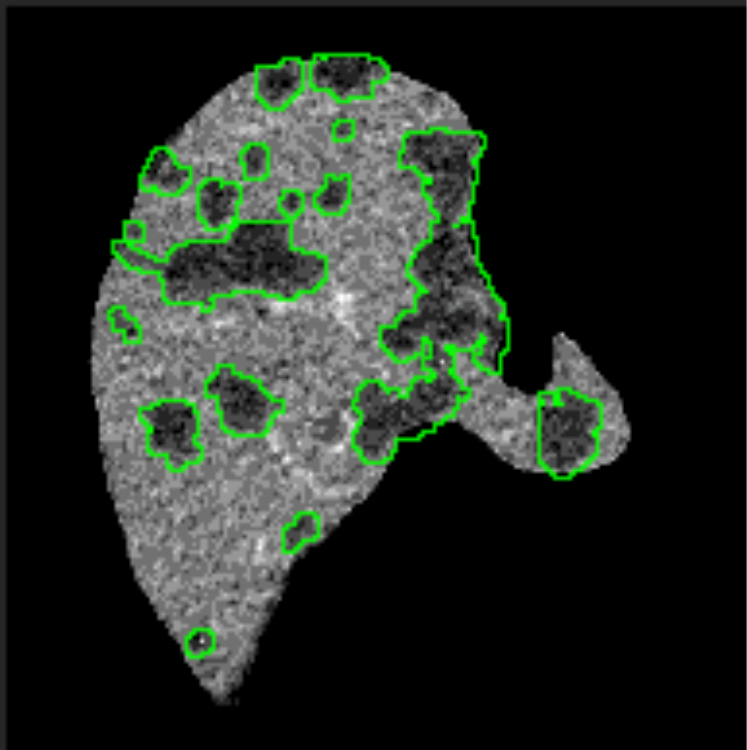}
\caption{Comparison between our predictions (green) and the LiTS annotations (blue) using the two-fold cross-validation on the training set. The red arrows indicate the artifacts which share similar intensity distribution to tumors annotated on the same slice but are missing in the annotations while being predicted by our method.}
\label{fig:comparison_gt_and_pred}
\end{figure}

\noindent
{\bf Experimental Results.}
Since we do not have the ground truth for the LiTS test set, we first perform the two-fold cross-validation on the training set to quantitatively and qualitatively evaluate the performance of our proposed method. We obtain $67.3\%$ Dice per case and some visual samples are presented in Fig.~\ref{fig:comparison_gt_and_pred}. 

\begin{figure}[t!]
\vspace{-0.1in}
\centering
\begin{tikzpicture}[thick, spy using outlines={circle,lens={scale=2}, width=1.5cm, height=1.5cm, connect spies}]
 \node (reg_id2)
 {\includegraphics[width=1.0\textwidth]{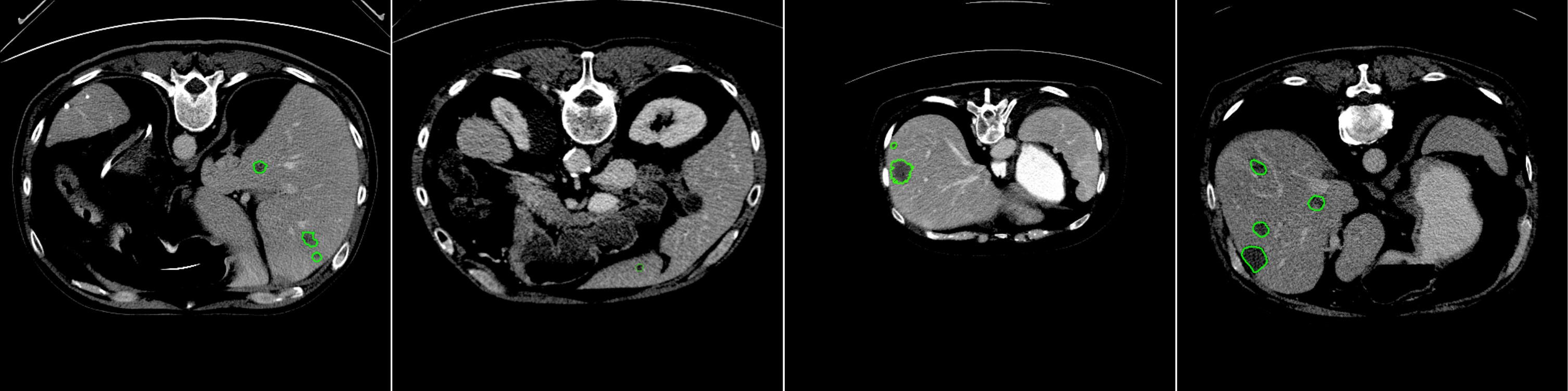}};
 \spy [yellow] on (-1.1, -0.5) in node (tspy) [left,fill=black!5] at (-1.6, -0.75);
 \spy [yellow] on (0.85, 0.3) in node (tspy) [left,fill=black!5] at (1.1, -0.75);
\end{tikzpicture}  
\caption{Examples of our predictions (indicated by the green lines) on the LiTS test set. }
\label{fig:test_result}
\end{figure}

By evaluating on the test set of the LiTS challenge, we list in Table~\ref{tab:result} our result with comparison to those of the currently published approaches. According to the Dice score per case, the most important metric for measuring an algorithm's performance on the LiTS challenge, our approach is at the second rank, following after the H-Dense U-Net~\cite{li2018h}, which however needs pre-training. In addition, our small lesion segmentation could be an add-on component to their method for a further improvement in segmenting small tumors as mentioned in their future work. Our experiment shows that the dice per case can be improved from 0.61 to 0.681 with the consideration of the small lesion segmentation. Worth to mention that, among the methods without pre-training and post-processing, our method has the best performance in terms of the dice per case score. Figure~\ref{fig:test_result} reports some visual results of our predictions on the LiTS test set.

\noindent
{\bf Observations on LiTS Annotations.} As shown in Fig.~\ref{fig:comparison_gt_and_pred}, the LiTS annotations have both over-segmentation and under-segmentation issues. For the large lesions in both cases shown in Fig.~\ref{fig:comparison_gt_and_pred}, our predictions better fit lesions compared to the ground truth; while for the small lesions, our predictions locate more lesions potentially missing in the ground truth. Similar observations have been reported in~\cite{chlebus2018automatic}. Due to the imperfect ground truth provided by the LiTS challenge, we argue that the metrics computed against the ground truth probably could not be the only way to compare the segmentation results. Visual results could be considered as well and ours indicate the efficiency of our method.  





\begin{table}[t]
\centering
\begin{tabular}{l|c|c|c|c}
\hline
Method &  Dice Per Case & Dice Overall & \makecell{No \\Pre-training} & \makecell{No\\Post-processing} \\ 
\hline
H-Dense U-Net \cite{li2018h} &  0.722 & 0.824 &  & \checkmark \\
Multiple U-Nets \cite{chlebus2018automatic} & 0.680 & 0.796 & \checkmark & \\
2.5 D U-Net \cite{han2017automatic} &  0.670  & --- & \checkmark & \\
CDNN~\cite{yuan2017hierarchical}& 0.657 & 0.820 & \checkmark & \checkmark\\
FED-Net \cite{chen2019feature} & 0.650 & 0.766 & \checkmark & \\
AH-Net \cite{liu20183d} & 0.634 & 0.834 & \checkmark & \checkmark \\
RA U-Net \cite{jin2018ra} & 0.595 & 0.795 & \checkmark & \checkmark \\
BS-Unet \cite{li2018bottleneck} & 0.552 & 0.729 & \checkmark & \checkmark \\
\hline
Ours (w/o small lesion) & 0.681 (0.610) & 0.813 (0.776) & \checkmark & \checkmark \\
\hline
\end{tabular}
\caption{Comparison among published approaches and ours on the LiTS challenge.}
\label{tab:result}
\end{table}

\section{Discussion and Conclusion}
\label{sec:discussion}

In this paper we proposed a hybrid 2D and 3D neural network for segmenting the liver lesions. Especially, we designed a dedicated 3D segmentation network for the small lesions in the liver. This 3D segmentation network could be an add on to a network that suffers from segmenting small objects. Moreover, we observed the imperfect annotations provided by the LiTS data set, which hinders the further improvement of a network's learning performance and makes the evaluation results questionable for comparison. How to handle and fully leverage such imperfect labels will be our future work. We also plan to extend our framework to other medical applications with the lesion/tumor or small object segmentation.
\section{Acknowledgement}
This work has been supported by NSF 1755970.


\bibliographystyle{splncs04}
\bibliography{main.bib}

\end{document}